\documentclass[pra,twocolumn,showpacs,superscriptaddress,10pt]{revtex4-1}
\usepackage[colorlinks,citecolor=blue,linkcolor=red,dvipdfm]{hyperref}
\usepackage{amsmath,amssymb,graphicx}
\usepackage{times}

\begin{document}

\title{Gaussian-like and flat-top solitons of atoms with spatially modulated repulsive interactions}

\author{Liangwei Zeng}
\affiliation{State Key Laboratory of Transient Optics and Photonics, Xi'an
Institute of Optics and Precision Mechanics of Chinese Academy of Sciences, Xi'an 710119, China}
\affiliation{University of Chinese Academy of Sciences, Beijing 100049, China}

\author{Jianhua Zeng}
\email{\underline{zengjh@opt.ac.cn}}
\affiliation{State Key Laboratory of Transient Optics and Photonics, Xi'an
Institute of Optics and Precision Mechanics of Chinese Academy of Sciences, Xi'an 710119, China}
\affiliation{University of Chinese Academy of Sciences, Beijing 100049, China}

\begin{abstract}
Solitons, nonlinear particle-like excitations with inalterable properties (amplitude, shape, and velocity) as they propagate, are omnipresent in many branches of science---and in physics in particular. Flat-top solitons are a novel type of bright solitons that have not been well explored in pure nonlinear media. Here, a model of nonlinear Kerr (cubic) media of ultracold atoms with spatially modulated repulsive interactions is proposed and shown to support a vast variety of stable flat-top matter-wave solitons, including one-dimensional (1D) flat-top fundamental and multipole solitons, two-dimensional (2D) flat-top fundamental and vortex solitons. We demonstrate that by varying the relevant physical parameters (nonlinearity coefficient and chemical potential) the ordinary bright (gaussian) solitons can transform into the novel flat-top solitons. The (in-)stability domains of the flat-top soliton families are checked by means of linear stability analysis and reconfirmed by direct numerical simulations. This model is generic in the contexts of nonlinear optics and Bose-Einstein condensates, which provide direct experimental access to observe the predicted solutions.
\end{abstract}

\maketitle

\section{Introduction}
Solitons (solitary waves), nonlinear excitations that preserve their shape and velocity during propagation, play a vital role for understanding the wave-matter interactions in diverse fields spreading from water waves to plasma waves, and from light waves in optics to matter waves in condensed matters \cite{S1,S2,S3,S4,REV1,REV2,REV3,REV4,REV5}.  Soliton studies are attractive to basic scientists, and to applied scientists with an interest in constantly innovating soliton-related technologies, e.g., the  soliton-driven ultrahigh-speed communications have revolutionized the conventional optical-transmission technologies \cite{SOC,SOF}.

The soliton phenomena are omnipresent in nature arising from an exquisite balancing mechanism between the wave dispersion (or diffraction) property and nonlinearity of the medium, of which a focusing nonlinearity in uniform media is responsible for generating bright solitons whose effective mass is positive \cite{PPS0,PPS1}. Recent studies demonstrated that bright gap solitons, a new type of solitons existing in nonlinear periodic systems \cite{PPS1,PPS3,PPS4}, also existed in defocusing nonlinearity, assisted by linear periodic potentials---photonic crystals \cite{BGS,WGA1,WGA2,WGA3} and lattices \cite{PTL0,PTL2} in optics, and optical lattices in Bose-Einstein condensates (BECs) \cite{PTL4,TPOL,EPOL,TPOL2,EPFTL2,EPFTL3}---by which the negative effective mass of the gap solitons was made possible. Bright solitons can also be supported by nonlinear lattices whose nonlinearity landscapes change periodically in magnitude and even the sign \cite{NL}. Despite the nonlinear lattices could readily support one-dimensional (1D) stable bright solitons, using them to stabilizing higher-dimensional solitons is still a challenge. Purely nonlinear lattices \cite{NL,NL1,NL2,NL3,NL3,NL4,NL5,NL6} and the combined linear-nonlinear lattices \cite{LNL1,LNL2} are now widely used to support bright solitons which, however, do not exist in  localized and periodic defocusing nonlinearities.

In contrast to the common beliefs, recent studies demonstrated that a pure nonlinear medium with spatially inhomogeneous defocusing nonlinearity whose local strength grows fast enough toward the periphery \cite{DFNBS2,DFNBS3,DFNBS4,DFNBS5,DFNBS6,DFNBS7,DFNBS8,DFNBS9,DFNBS10}, provides a fertile ground for the generation and stabilization of bright solitons and other localized modes in all three dimensions, including solitary vortices (with arbitrarily vorticity) and vortex rings \cite{AVS}, skyrmions \cite{SKYRM}, hopfions, soliton gyroscopes \cite{GYRO}, complex hybrid modes, localized dark solitons and vortices \cite{LVS}, to name a few of them.

Flat-top solitons---the solitary waves with constant intensity---are a new type of bright solitons that have been less understood in physics. Flat-top solitons appear as localized coherent structures in several areas of physics from nonlinear optics \cite{FTS1,FTS2} to fluid physics \cite{FTSFD} and plasma physics\cite{FTSPP}. In terms of nonlinear optics, specifically, Flat-top solitons exist in cubic-quintic nonlinear media \cite{FTSCQ1,FTSCQ2}, cubic-nonlinear media with non-Hermitian potentials \cite{FTSV}, and nonlinear metamaterials \cite{FTSMM}. It is commonly believed that stable flat-top solitons can only exist in cubic-quintic nonlinear media or with the help of linear potentials \cite{FTSBEC}, since they are formed from an elegant balance between defocusing quintic nonlinearity or linear potential and self-focusing cubic nonlinearity. While flat-top solitons exist in cubic nonlinear media their stability is still in debate \cite{FTSUNS} which, therefore, is a critical issue to be clarified. Very recently, we further predicted that the flat-top solitons can be stable profiles in purely Kerr media \cite{FTSPN}. However, the nonlinearity landscapes of such purely Kerr media are not smooth and thus may be difficult to realize in experiment. Besides, the transitions from ordinary Gaussian-like solitons and vortexes to their flat-top counterparts in the purely defocusing media are waiting for clear illustration.

In this paper, we discover that flat-top solitons and vortices can be supported by spatially inhomogeneous defocusing nonlinearities with simple forms, which may be implemented in experiments. Importantly, by varying the chemical potential (propagation constant in nonlinear optics) and nonlinearity parameter of the considered physical setting, the transitions from ordinary Gaussian-like solitons and vortexes to flat-top ones are also clearly clarified. The rest of this article is organized as follows. The theoretical model and linear stability analysis are introduced in Secs. \ref{sec2a} and \ref{sec2b} respectively. In Sec. \ref{sec3a}, we report the 1D numerical results, including the transitions from Gaussian-like solitons to flat-top counterparts for fundamental, dipole and multipole solitons. The 1D fundamental, dipole and tripole flat-top solitons are demonstrated to be completely stable. In Sec. \ref{sec3b}, we present the 2D numerical results, including the transitions from Gaussian-like solitons and vortices to flat-top ones. The 2D flat-top vortex solitons with topological charge $m\leq1$ are all stable. Finally, we conclude this paper in Sec. \ref{sec4}.

\section{Theoretical model and linear stability analysis}
\subsection{Theoretical model and corresponding solutions}
\label{sec2a}
We consider the evolution of a matter-wave packet in an atomic BEC cloud (medium) with spatially inhomogeneous defocusing nonlinearity whose physical model is in the framework of the Gross-Pitaevskii equation (or nonlinear Schr\"{o}dinger equation) for the mean-field wave function $\psi$, written in the dimensionless form:

\begin{equation}
i\frac{\partial \psi}{\partial t}=-\frac{1}{2}\nabla^2\psi+\mathrm{g}(r)\left|\psi\right|^2\psi.
\label{NLSE}
\end{equation}
Where $r=(x,y)$ is the set of transverse coordinates and Laplacian $\nabla ^{2}=\partial
_{x}^{2}+\partial _{y}^{2}$. Regarding the light beam propagation in optical media for the field amplitude $\psi$, the time $t$ should be replaced by propagation distance $z$. The defocusing nonlinear strength ${\rm g}(r)>0$ is spatially inhomogeneous modulated and in the form of:
\begin{equation}
\mathrm{g} (r)=\left\{
\begin{array}{c}
\mathrm{g}_{0},r\leq r_{0}, \\
\mathrm{g}_{0}+(r-r_{0})^\alpha,r>r_{0},
\end{array}
\right.
\label{gprofile}
\end{equation}
here $\mathrm{g}_{0}$ and $\alpha$ are positive constants. In 1D case, $r$ and $r_{0}$ should be replaced by $|x|$ and $x_{0}$, respectively. Notably, when $\mathrm{g}_{0}=1$ and $r_0=0$,  Eq. (\ref{gprofile}) reduces to the model in Ref. \cite{DFNBS2} where the nonlinearity yields $\mathrm{g}(r)=1+|r|^\alpha$ and which, however, could not readily support the flat-top solitons, since the lacking of a flat region in the center.

With real chemical potential $\mu$ in atomic condensate, the stationary solutions are searched by $\psi=\phi~{\rm exp}(-i\mu t)$, the resulting stationary wave function should be met:
\begin{equation}
\mu \phi=-\frac{1}{2}\nabla^2\phi+\mathrm{g}(r)\left|\phi\right|^2\phi.
\label{NLSES}
\end{equation}
In optical field, chemical potential $\mu$ should be replaced by the propagation constant $-b$.

To judge whether a soliton is flat-top or not, we define a criterion as the parameter, $\Gamma=r_F/r_H$, here $r_F$ denotes the top width of the solitons and $r_H$ being the half-peak width of solitons; specifically, we call the flat-top solitons if $\Gamma\geq0.5$.

\subsection{Linear stability analysis}
\label{sec2b}
To settle the stability of the so-found flat-top solutions, it is indispensable to scrutinize the relevant linear stability analysis. For this, we set the perturbed 1D wave function as $\psi=[\phi(x)+p(x){\rm exp}(\lambda t)+q^*(x){\rm exp}(\lambda ^*t)]{\rm exp}(-i\mu t)$, where $\phi(x)$ is undisturbed wave function, $p(x)$ and $q^*(x)$ are small perturbations. Substituting it into Eq. (\ref{NLSE}) and after linearization, we obtain the resulting eigenvalue problem as:

\begin{equation}
\left\{
\begin{aligned}
i\lambda p=-\frac{1}{2}\nabla^2p-\mu p+\mathrm{g}\phi^2(2p+q),\\
i\lambda q=+\frac{1}{2}\nabla^2q+\mu q-\mathrm{g}\phi^2(2q+p).
\end{aligned}
\label{LAS1D}
\right.
\end{equation}

For 2D circumstance, the perturbed wave function is defined by $\psi=[\phi(r)+p(r){\rm exp}(in\theta+\lambda t)+q(r){\rm exp}(-in\theta+\lambda^* t)]{\rm exp}(im\theta-i\mu t)$, which yields the eigenvalue problem:

\begin{equation}
\left\{
\begin{aligned}
i\lambda p=-\frac{1}{2}\left[\nabla^2+\frac{1}{r}\nabla-\frac{(m+n)^2}{r^2}\right]p-\mu p+\mathrm{g}\phi^2(2p+q),\\
i\lambda q=+\frac{1}{2}\left[\nabla^2+\frac{1}{r}\nabla-\frac{(m-n)^2}{r^2}\right]q+\mu q-\mathrm{g}\phi^2(2q+p).
\end{aligned}
\label{LAS2D}
\right.
\end{equation}

It follows from the eigenvalue problems (\ref{LAS1D}) and (\ref{LAS2D}) that the perturbed solitonic solutions are stable only when the real part of all the eigenvalues ($\lambda$) is null, that is, ${\rm Re}(\lambda)=0$. The stationary solutions were found by the Newton's method; their stability was subsequently investigated by computing the eigenvalue equations (\ref{LAS1D}) and (\ref{LAS2D}), and was finally rechecked by direct simulations of the thus-found solutions under small perturbations in the evolution equation (\ref{NLSE}).

\begin{figure}[tbp]
\begin{center}
\includegraphics[width=1\columnwidth]{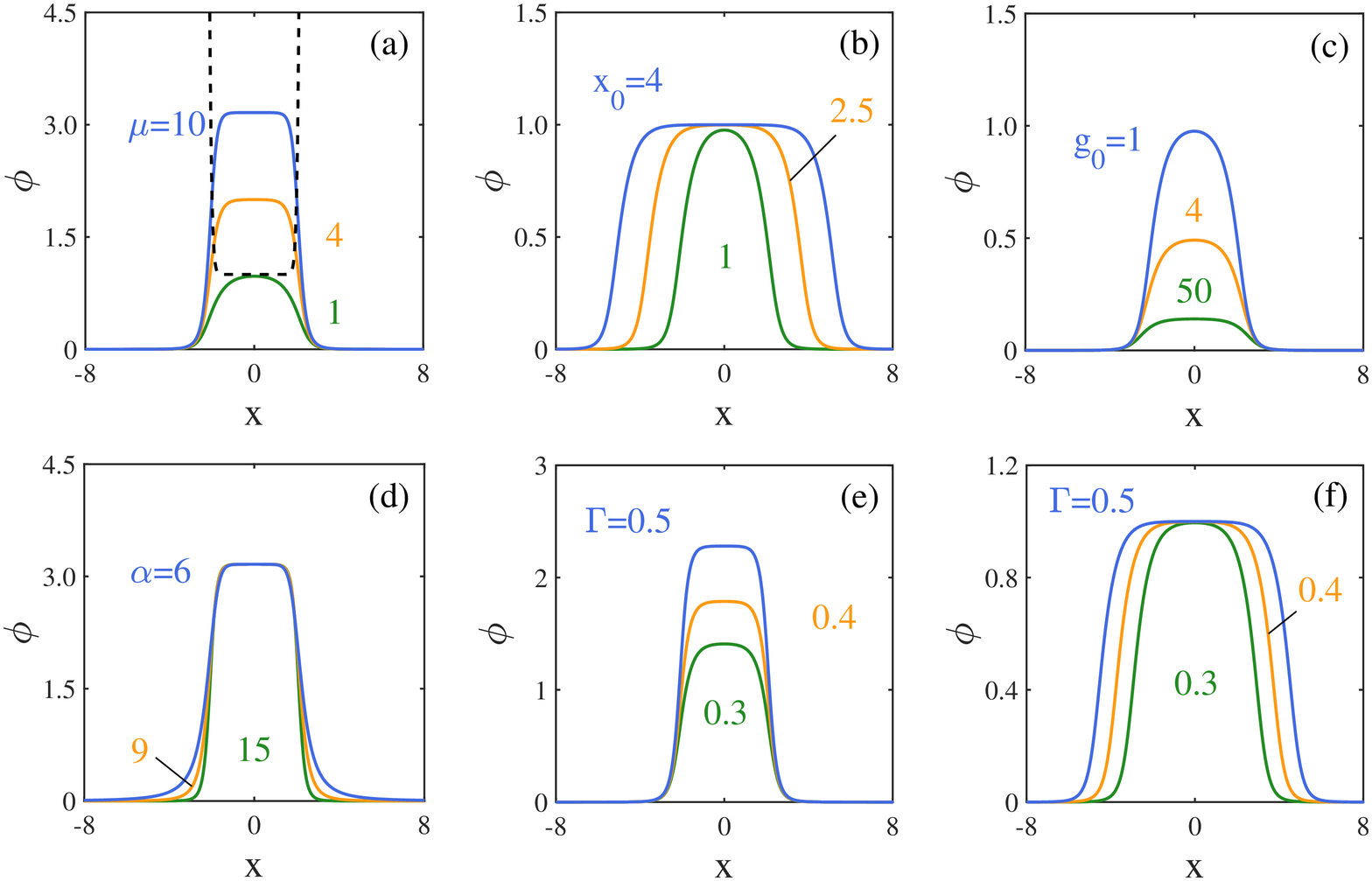}
\end{center}
\caption{Profiles of 1D fundamental solitons: (a) with different values of chemical potential $\mu$ at $\mathrm{g}_0=1$, $x_0=1$, $\alpha=12$; (b) with different values of $x_0$ at $\mathrm{g}_0=1$, $\alpha=12$, $\mu=1$; (c) with different values of $\mathrm{g}_0$ at $x_0=1$, $\alpha=12$, $\mu=1$; (d) with different values of $\alpha$ at $\mathrm{g}_0=1$, $x_0=1$, $\mu=10$; (e) with different $\Gamma$ (varying $\mu$) at $\mathrm{g}_0=1$, $x_0=1$, $\alpha=12$; (f) with different $\Gamma$ (varying $x_0$) at $\mathrm{g}_0=1$, $\alpha=12$, $\mu=1$. Here and in Figs. \ref{fig2} and \ref{fig5} below, black dashed lines denote the nonlinearity profiles given by Eq. (\ref{gprofile}).}
\label{fig1}
\end{figure}

\section{Numerical results}
\subsection{1D fundamental and multipole flat-top solitons}
\label{sec3a}
We firstly report 1D numerical solutions of the present model based on the nonlinearity modulation profile (\ref{gprofile}). Various kinds of 1D localized modes of flat-top types are obtained, including fundamental solitons and multipole ones with different numbers of nodes $k$. By changing the relevant physical parameters, e.g., nonlinearity parameters ($x_0$, $\mathrm{g}_0$) and chemical potential $\mu$, the ordinary bright solitons may evolve into flat-top modes, as exemplified in Figs. \ref{fig1}(a$\sim$c), from which one can clearly see that the 1D fundamental bright solitons become flatter with the increase of parameters $x_0$, $\mathrm{g}_0$ and $\mu$, proving that such new type of flat-top solitons is robust and controllable. Depicted in Fig. \ref{fig1}(d) is a collection of fundamental flat-top solitons at different $\alpha$. Note that $\alpha$ is not a critical parameter to judge whether a soliton is flat-top or not, despite it can greatly affect the decaying tails of the solitons. Figs. \ref{fig1}(e,f) show that the solitons become flat-top shape when $\Gamma=0.5$, and the solitons are still not flat enough when $\Gamma<0.5$.

\begin{figure}[tbp]
\begin{center}
\includegraphics[width=1\columnwidth]{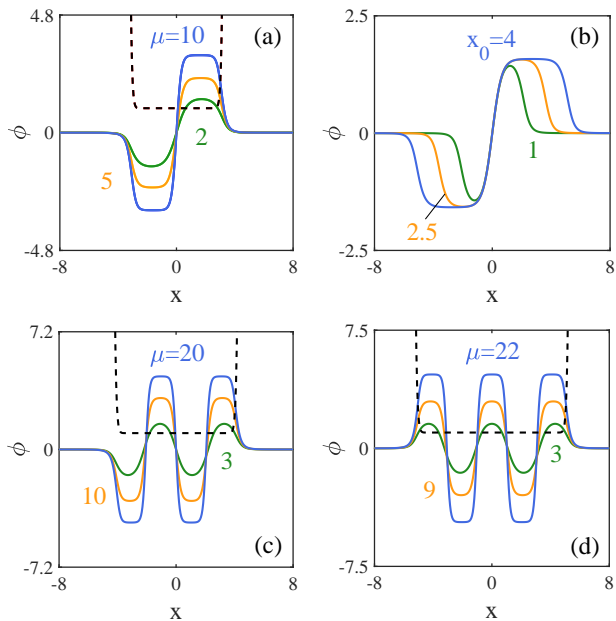}
\end{center}
\caption{Profiles of 1D dipole or multipole solitons: (a) dipole ($k=1$) solitons with different values of $\mu$ at $\mathrm{g}_0=1$, $x_0=2$; (b) dipole solitons with different values of $x_0$ at $\mathrm{g}_0=1$, $\mu=2.5$; (c) multipole ($k=3$) solitons with different values of $\mu$ at $\mathrm{g}_0=1$, $x_0=3$; (d) multipole ($k=4$) solitons with different values of $\mu$ at $\mathrm{g}_0=1$, $x_0=4$. This and other figures below are displayed at $\alpha=12$.}
\label{fig2}
\end{figure}

\begin{figure}[tbp]
\begin{center}
\includegraphics[width=1\columnwidth]{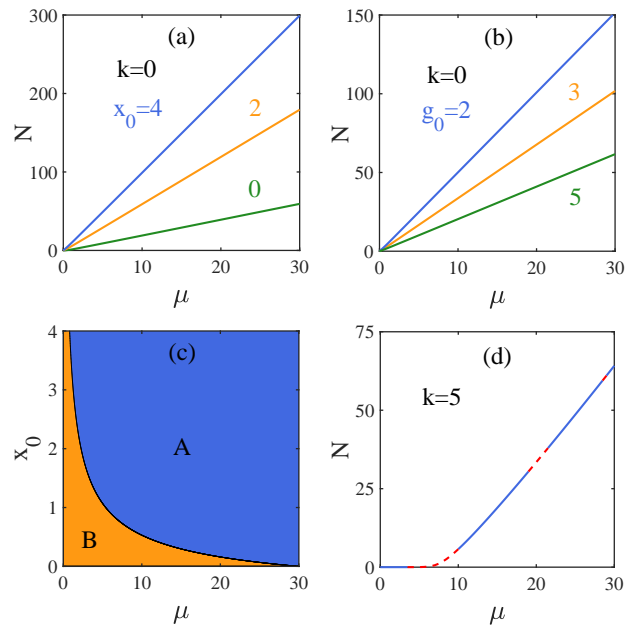}
\end{center}
\caption{Number of atoms $N$ versus $\mu$ for 1D fundamental solitons: (a) with different $x_0$ at $\mathrm{g}_0=1$; (b) with different $\mathrm{g}_0$ at $x_0=4$. (c) The flat-top (blue) and ordinary (yellow) modes domains for 1D fundamental solitons in the $(\mu,x_0)$ plane corresponding to $\mathrm{g}_0=1$. (d) $N$ versus $\mu$ for 1D multipole solitons with $k=5$ at $\mathrm{g}_0=1$, $x_0=1$; the blue solid and red dashed lines represent the stable and unstable domains respectively.}
\label{fig3}
\end{figure}

Besides the fundamental flat-top modes, 1D high-order modes with different numbers of nodes (zeros of the wave function) $k$ can also be supported by the current nonlinearity profile given by Eq. (\ref{gprofile}) and, markedly, they are robust. Shown in Figs. \ref{fig2}(a,b) are the 1D dipole ($k=1$) modes at different values of $\mu$ and $x_0$, which verify that the larger these parameters are, the flatter the dipole solitons are, in analogous to their fundamental counterparts in Fig. \ref{fig1}. This feature keeps also for 1D high-order solitons, as seen from the Figs. \ref{fig2}(c,d) for multipole solitons with $k=3,4$.

The dependence of number of atoms $N$ to the chemical potential $\mu$ for the fundamental solitons, with different values of $x_0$ and $\mathrm{g}_0$, is respectively displayed in Figs. \ref{fig3}(a,b), which show the linear dependence $N (\mu)$. Our study reveals that the $N$ grows and the shapes of solitons get flatter with an increase of $\mu$ and $x_0$ for all types of 1D solitons (fundamental and multipole ones).

\begin{figure}[tbp]
\begin{center}
\includegraphics[width=1\columnwidth]{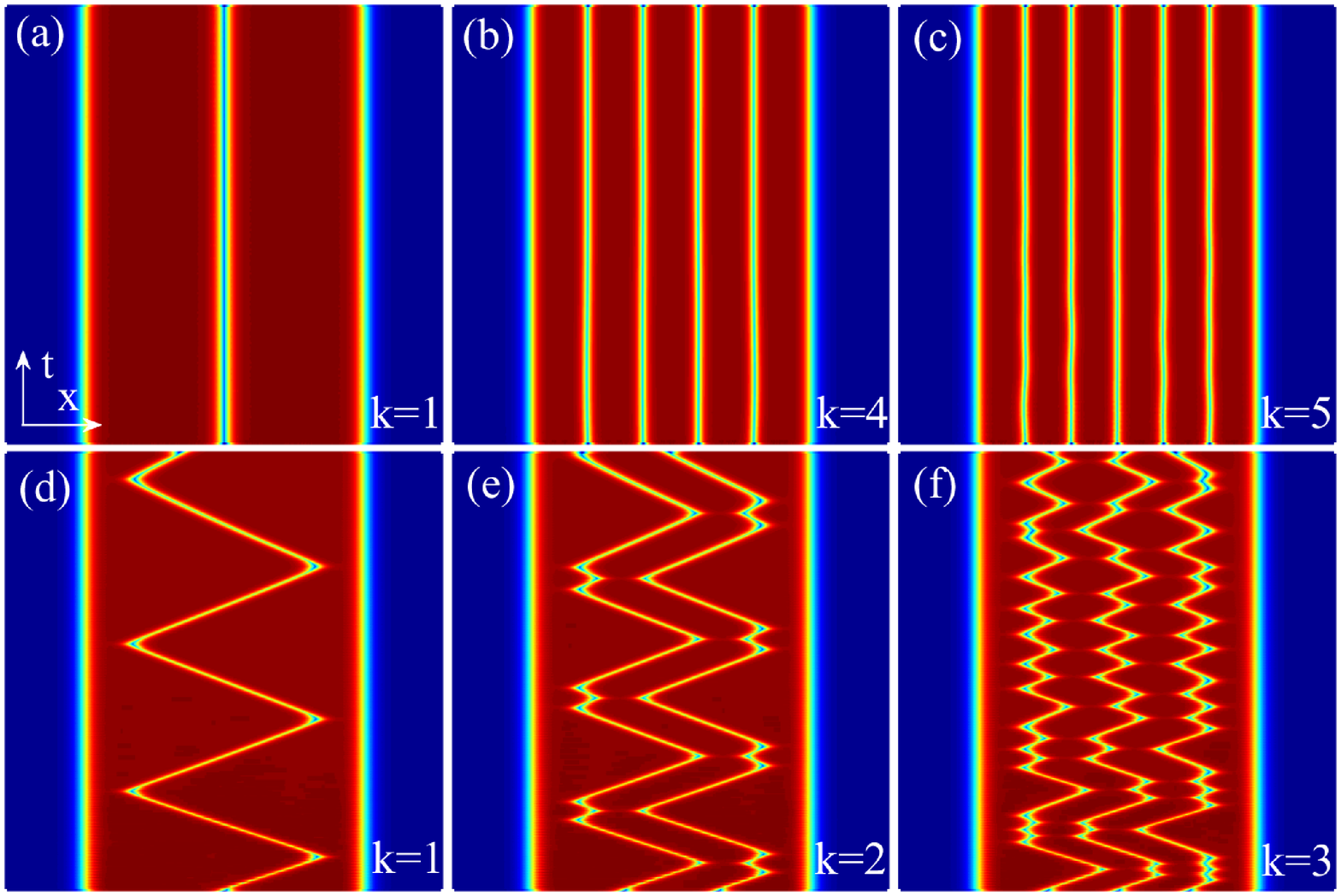}
\end{center}
\caption{Stable evolution of 1D solitons at $\mathrm{g}_0=1$, $x_0=4$: (a) dipole soliton at $\mu=9$; (b) multipole soliton with $k=4$ at $\mu=21$; (c) multipole soliton with $k=5$ at $\mu=25$. Oscillations of 1D solitons at $\mathrm{g}_0=1$, $x_0=4$, $\mu=8$, phase tilt $\omega=0.2$: (d) dipole soliton; (e) tripole ($k=2$) soliton; (f) multipole soliton with $k=3$.}
\label{fig4}
\end{figure}

The transition from ordinary solitons to flat-top ones is a key issue, since flat-top solitons have been used for high-order-harmonic generation in experiments \cite{FTS2}. the flat-top (blue) and ordinary (yellow) modes domains for 1D fundamental solitons in the $(\mu,x_0)$ plane is shown in Fig. \ref{fig3}(c). The linear stability analysis depended on eigenvalue problems (\ref{LAS1D}) and the direct simulations of Eq. (\ref{NLSE}) demonstrate that the 1D solitons with $k=0,1,2$ are robustly stable with $\mu$ at least up to $30$. The result of stability area for 1D multipole solitons with $k=5$ is depicted in Fig. \ref{fig3}(d) where blue solid and red dashed lines denote stable and unstable domains. To further investigate the dynamical properties of such 1D multipole modes, we show in Figs. \ref{fig4} (a$\sim$c) the stable evolutions of perturbed solitons with different $k$. It is seen from the figures that both 1D dipole solitons and high-order (with $k=4,5$) ones keep their coherence over long time evolution ($t=10^3$).

It is well known that the definition of solitons in physics contains several preserved properties, the intrinsic coherence (of amplitude, shape, and velocity) during the movement, and quasielastic collisions. By multiplying with ${\rm exp}(i\omega X)$, where $\omega$ denotes a phase tilt, the solitons might begin to move. The evolutions of 1D multipole solitons ($k=1,2,3$) with a small kick $\omega$ is displayed in Figs. \ref{fig4} (d$\sim$f), which show that the solitons with $k=1,2$ start and maintain regular oscillations during the evolutions up to $t=10^3$, while irregular oscillations happen for the solitons with $k\geq3$.

\subsection{2D flat-top solitons and vortexes}
\label{sec3b}
\begin{figure}[tbp]
\begin{center}
\includegraphics[width=1\columnwidth]{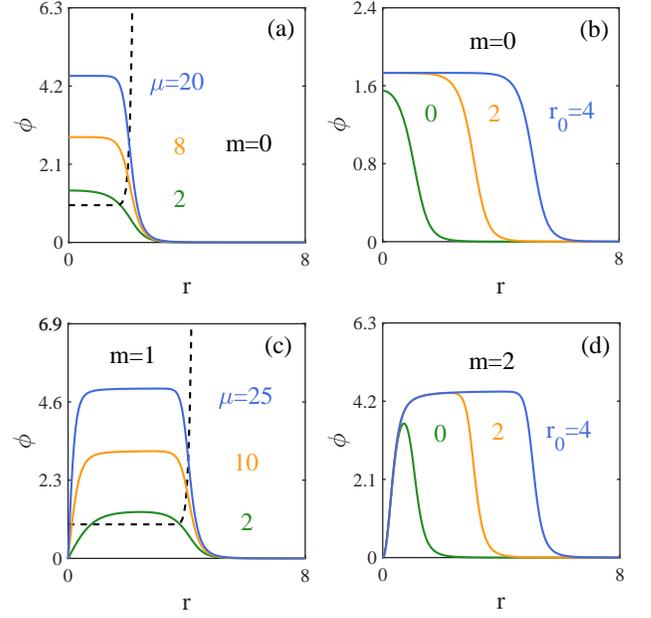}
\end{center}
\caption{Profiles of 2D solitons: (a) fundamental soliton with different values of chemical potential $\mu$ at $\mathrm{g}_0=1$, $x_0=1$; (b) fundamental soliton with different values of $r_0$ at $\mathrm{g}_0=1$, $\mu=3$; (c) vortex soliton with vortex charge $m=1$ for different values of $\mu$ at $\mathrm{g}_0=1$, $r_0=3$; (d) vortex soliton with vortex charge $m=2$ for different values of $r_0$ at $\mathrm{g}_0=1$, $\mu=20$.}
\label{fig5}
\end{figure}

This section reports numerical solutions of various 2D flat-top solitons and vortical ones and their stability domains are identified by virtue of linear stability analysis and direct simulations. Characteristic examples of 2D fundamental flat-top solitons are shown in Figs. \ref{fig5}(a,b). It is observed, obviously, that, the 2D fundamental solitons once again transform into flat-top ones if we increase gradually the values of $\mu$ and $r_0$, resembling their 1D counterparts in Fig. \ref{fig1}. In addition to the fundamental modes, our unique model given by nonlinearity profile (\ref{gprofile}) also give rise to a vast variety of flat-top high-order modes---vortex solitons, typical profiles of them are illustrated in Figs. \ref{fig5}(c) and \ref{fig5}(d), for vorticities (vortex charges) $m=1$ and $m=2$, respectively.

\begin{figure}[tbp]
\begin{center}
\includegraphics[width=1\columnwidth]{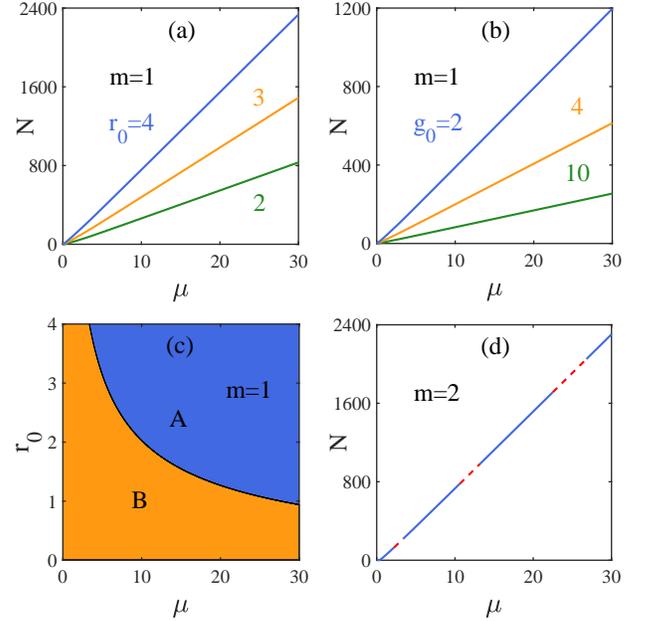}
\end{center}
\caption{Number of atoms $N$ versus $\mu$ for 2D vortex solitons with $m=1$: (a) with different $r_0$ at $\mathrm{g}_0=1$; (b) with different $\mathrm{g}_0$ at $r_0=4$. (c) The flat-top (blue) and ordinary (yellow) modes domains for 2D vortex solitons with $m=1$ in the $(\mu,r_0)$ plane corresponding to $\mathrm{g}_0=1$. (d) $N$ versus $\mu$ of 2D vortex solitons with $m=2$, $\mathrm{g}_0=1$, $r_0=4$; the blue solid and red dashed lines represent the stable and unstable domains respectively.}
\label{fig6}
\end{figure}

The relation between $N$ and $\mu$ for the 2D vortex solitons at $m=1$ with different values of $r_0$ and $\mathrm{g}_0$ is plotted in Fig. \ref{fig6}(a,b), demonstrating that the number of atoms $N$ increases linearly with $\mu$, similar to that for their 1D cases in Fig. \ref{fig3}(a,b). Note also that $N$ can be very large provided that the width $r_0$ is appropriate, e.g., $r_0=4$, making their (flat-top solitons and vortices) observations are more feasible in experiments of BEC. We show the flat-top and ordinary modes domains for 2D vortex solitons with $m=1$ in the $(\mu,r_0)$ plane in Fig. \ref{fig6}(c), where they are portrayed in blue and yellow respectively. The transition criterion used here for 2D flat-top solitons is the same as its 1D counterpart.

\begin{figure}[tbp]
\begin{center}
\includegraphics[width=1\columnwidth]{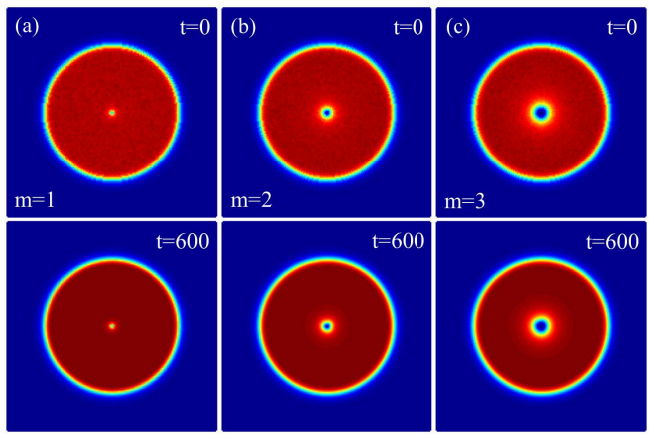}
\end{center}
\caption{Stable evolutions of 2D perturbed vortex solitons for $\mathrm{g}_0=1$, $r_0=4$: (a) with $m=1$ at $\mu=30$; (b) with $m=2$ at $\mu=18$; (c) with $m=3$ at $\mu=14$.}
\label{fig7}
\end{figure}

\begin{figure}[tbp]
\begin{center}
\includegraphics[width=1\columnwidth]{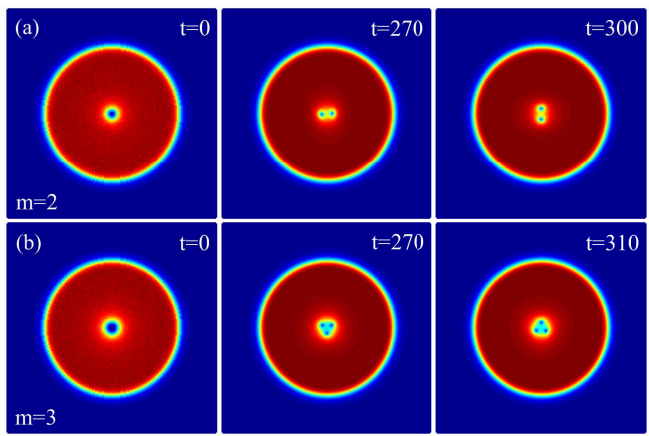}
\end{center}
\caption{Unstable evolutions of 2D perturbed vortex solitons for $\mathrm{g}_0=1$, $r_0=4$: (a) with $m=2$ at $\mu=12$; (b) with $m=3$ at $\mu=16$.}
\label{fig8}
\end{figure}

A nontrivial issue concerning both types of 2D flat-top localized modes---fundamental solitons and vortex ones---is their stability. For this, we analyze the relation between $N$ and $\mu$ for vortex solitons with $m=2$ in Fig. \ref{fig6}(d), which features alternating stability (solid) and instability (dashed) domains. We conclude that both the fundamental and vortex solitons with $m=1$ are always stable up to $\mu=30$. Systematic numerical simulations verify the stable evolutions of both localized modes, which keep their structures under the background of noise perturbation over time $t=10^3$, as displayed in Fig. \ref{fig7}. The unstable vortex solitons, with $m\geq2$, split into a set of multiple unitary vortices (with number equalling to $m$) that stand near the pivot, and rotating steadily around it, accounting for the conservation of angular momentum. Typical examples of these modes for $m=2,3$ are respectively portrayed in Figs. \ref{fig8}(a), \ref{fig8}(b).

\section{Discussion}
\label{sec4}

In this work we have theoretically and numerically investigated the existence and stability of a relatively new type of bright localized modes---flat-top solitons (and their vortex counterparts) in nonlinear cubic optical or matter-wave media, which are supported by spatially inhomogeneous defocusing nonlinearities whose local strength increases quickly enough from the center to periphery. Furthermore, we clearly show the transition from ordinary solitons to flat-top families by changing the chemical potential and the nonlinearity parameters. Such media with unique nonlinearity landscapes, beyond giving rise to varieties of fundamental flat-top solitons in one and two dimensions, also brings about families of higher-order modes, including the 1D solitons with different numbers of nodes $k$ (such as dipole and multipole states) and 2D flat-top vortex solitons with different vortex charges $m$. By use of linear stability analysis and direct numerical calculations of the evolutions of perturbed solutions, we confirm that the 1D solitons with number of nodes $k\leq2$ and 2D solitons with vorticity $m\leq1$ are exceptionally stable, while the existence of 1D solitons with $k\geq3$ and 2D solitons with $m\geq2$ is within alternating stability and instability areas.
Note that the 1D multipole solitons are similar to the linear modes of linear waveguides, however, they are essentially different, since they are formed by the purely nonlinear and linear media, respectively. Naturally, following the works \cite{DFNBS7,SKYRM,GYRO}, this paper can be extended to 3D setting, where rich physical dynamics can be explored widely.

\section{Acknowledgment}
This work was supported by the National Natural Science Foundation of China (grant numbers 61690222, 61690224), and partially by the Youth Innovation Promotion Association of the Chinese Academy of Sciences (grant number 2016357).

\end{document}